\documentstyle[11pt,lesiond,epsfig]{article}

\def\Journal#1#2#3#4{{#1} {\bf #2}, #3 (#4)}

\def\NPB{{\em Nucl. Phys.} B}
\def\PLB{{\em Phys. Lett.}  B}
\def\PRL{\em Phys. Rev. Lett.}
\def\PRD{{\em Phys. Rev.} D}

\def\kpnnp{$K^+ \to \pi^+ \nu\bar\nu$}
\def\bkpnnp{$B(K^+ \to \pi^+ \nu\bar\nu)$}
\def\kpnn0{$K_L\to\pi^0\nu\bar\nu$}
\def\bkpnn0{$B(K_L\to\pi^0\nu\bar\nu)$}
\def\kmm{$K_L\to\mu^+\mu^-$}
\def\bkmm{$B(K_L\to\mu^+\mu^-)$}
\def\kpp0{$K_L \to \pi^0 \pi^0$}

\def\be{\begin{equation}}
\def\ee{\end{equation}}
\def\bea{\begin{eqnarray}}
\def\eea{\end{eqnarray}}

\begin{document}
\begin{flushright}
{\bf BNL-67772}
\end{flushright}
\vspace*{4cm}

\title{RARE KAON DECAYS\footnote{Submitted to the proceedings
of the XXXVth Rencontres de Moriond session devoted to 
Electroweak Interactions and Unified Theories.}}

\author{ L. LITTENBERG }

\address{Physics Department, Brookhaven National Laboratory\\
Upton, NY 11973}

\maketitle\abstracts{
The current status of the study of rare kaon decays is reviewed.  Future
prospects are discussed.}

\newpage
\section{Introduction}

In recent years motivations for studying the rare decays of kaons have
been threefold.  First is the search for physics beyond the Standard
Model (SM).  Virtually every attempt to redress the theoretical
shortcomings of the SM predicts some degree of lepton flavor
violation (LFV).  This can be manifested in kaon decays such as $K_L
\to \mu^{\pm} e^{\mp}$.  Such decays have very good experimental
signatures and can therefore be pursued to remarkable sensitivities.
These sensitivities in turn correspond to extremely high energy scales
in models where the only suppression is that of the mass of the
exchanged field.

	There are also decays which are allowed by the SM, but which
are extremely suppressed.  In the the most interesting of these the
leading contribution is a G.I.M.-suppressed~\cite{GIM} one-loop process
quite sensitive to fundamental SM parameters such as $V_{td}$ and $m_t$.
These decays are also potentially very sensitive to new physics.

	Finally there are a number of long-distance-dominated decays which
serve as a testing ground for theoretical techniques such as chiral lagrangians
that seek to account for the low-energy behavior of QCD.  Knowledge of certain
of these decays is also needed to extract more fundamental information from
some of the one-loop processes.

	This field has been quite active in the last few years, so one
has to be selective in a short review.  This can be established by 
inspection of Table \ref{decays} which lists the
decays for which results have been announced in the last couple of years
and those that I happen to know are under analysis.

\begin{table}[h]
\caption{Rare $K$ decay modes under recent or on-going study.\label{decays}}
\vspace{0.4cm}
\begin{center}
\begin{tabular}{|c|c|c|c|}
\hline
&& & \\
$K^+ \to \pi^+ \nu\bar\nu$ & $K_L \to \pi^0 \nu\bar\nu$ &
$K_L \to \pi^0 \mu^+\mu^-$ & $K_L \to \pi^0 e^+e^-$ \\
$K^+ \to \pi^+ \mu^+\mu^-$ & $K^+ \to \pi^+ e^+e^-$ &
$K_L \to  \mu^+\mu^-$ & $K_L \to  e^+e^-$ \\
$K^+ \to \pi^+ e^+ e^- \gamma$ & $K^+ \to \pi^+ \pi^0 \nu\bar\nu$ &
$K_L \to e^{\pm} e^{\mp} \mu^{\pm} \mu^{\mp}$ & $K^+ \to \pi^+ \pi^0 \gamma$ \\
$K_L \to \pi^+ \pi^- \gamma$ & $K_L \to \pi^+ \pi^- e^+ e^-$ &
$K^+ \to \pi^+ \pi^0 e^+ e^-$ & $K^+ \to \pi^0 \mu^+ \nu \gamma$ \\
$K_L \to \pi^0 \gamma \gamma$ & $K^+ \to \pi^+ \gamma \gamma$ &
$K^+ \to \mu^+ \nu \gamma$ & $K^+ \to e^+ \nu e^+ e^-$ \\
$K^+ \to \mu^+ \nu e^+ e^-$ & $K^+ \to e^+ \nu \mu^+\mu^-$&
$K_L \to e^+ e^- \gamma$ & $K_L \to \mu^+ \mu^- \gamma$ \\
$K_L \to e^+ e^- \gamma\gamma$ & $K_L \to \mu^+ \mu^- \gamma\gamma$ &
$K_L \to e^+ e^- e^+ e^-$ & $K_L \to \pi^0 e^+ e^- \gamma$ \\
$K^+ \to \pi^+ \mu^+e^-$ & $K_L \to \pi^0 \mu^{\pm} e^{\mp}$ &
$K_L \to \mu^{\pm} e^{\mp}$ & $K^+ \to \pi^- \mu^+ e^+$ \\
$K^+ \to \pi^- e^+ e^+$ & $K^+ \to \pi^- \mu^+ \mu^+$ &
$K^+ \to \pi^+ X^0$ & $K_L \to e^{\pm} e^{\pm} \mu^{\mp} \mu^{\mp}$ \\ \hline
\end{tabular}
\end{center}
\end{table}

	For those who want a more extensive discussion of this subject,
a new review by Barker and Kettell is now available~\cite{steve}.

\section{Beyond the Standard Model}

	There were a series of dedicated $K$ decay experiments on the
subject of lepton flavor violation at the Brookhaven AGS during the
1980's and 90's.  These advanced the sensitivity to this sort of
phenomenon by many orders of magnitude.  In addition there were
``by-product'' results on this and other BSM topics from most of the
other $K$ experiments in business during this period.

\begin{figure}
\begin{minipage}[b]{.475\linewidth}
\psfig{figure=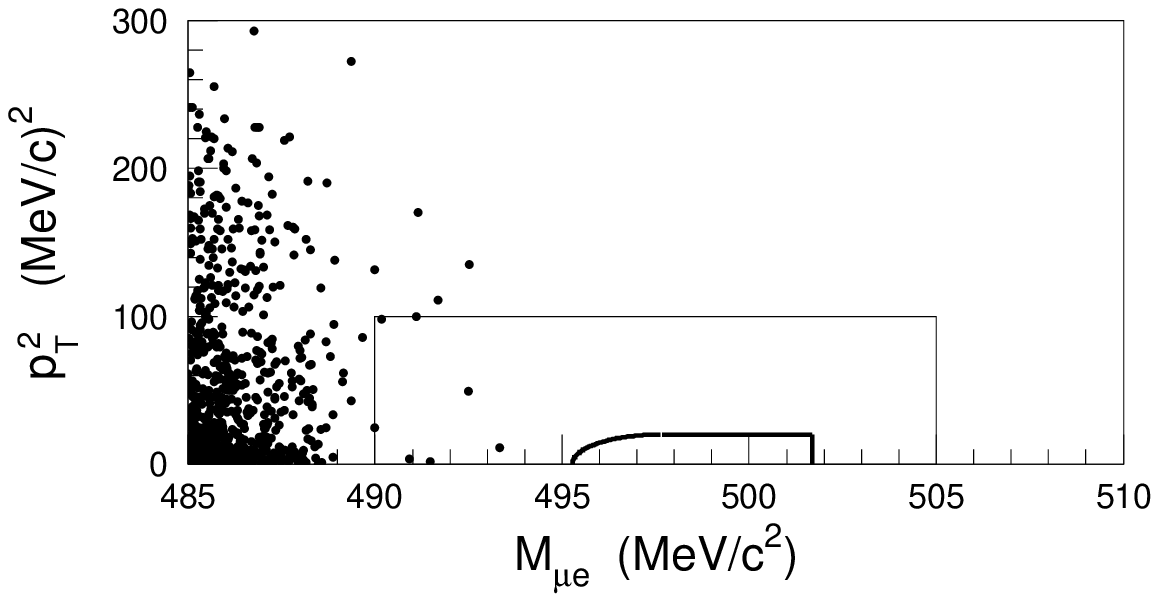,width=\linewidth}
\caption{Distribution of $m_{\mu e}$ vs $p_T^2$ for events passing all other 
cuts (AGS E871).  Rectangular contour bounds the ``blind'' region not 
interrogated until all cuts were finalized.  Inner contour bounds the signal 
region.
\label{fig:871mue}}
\end{minipage}\hfill
\begin{minipage}[b]{.445\linewidth}
\psfig{figure=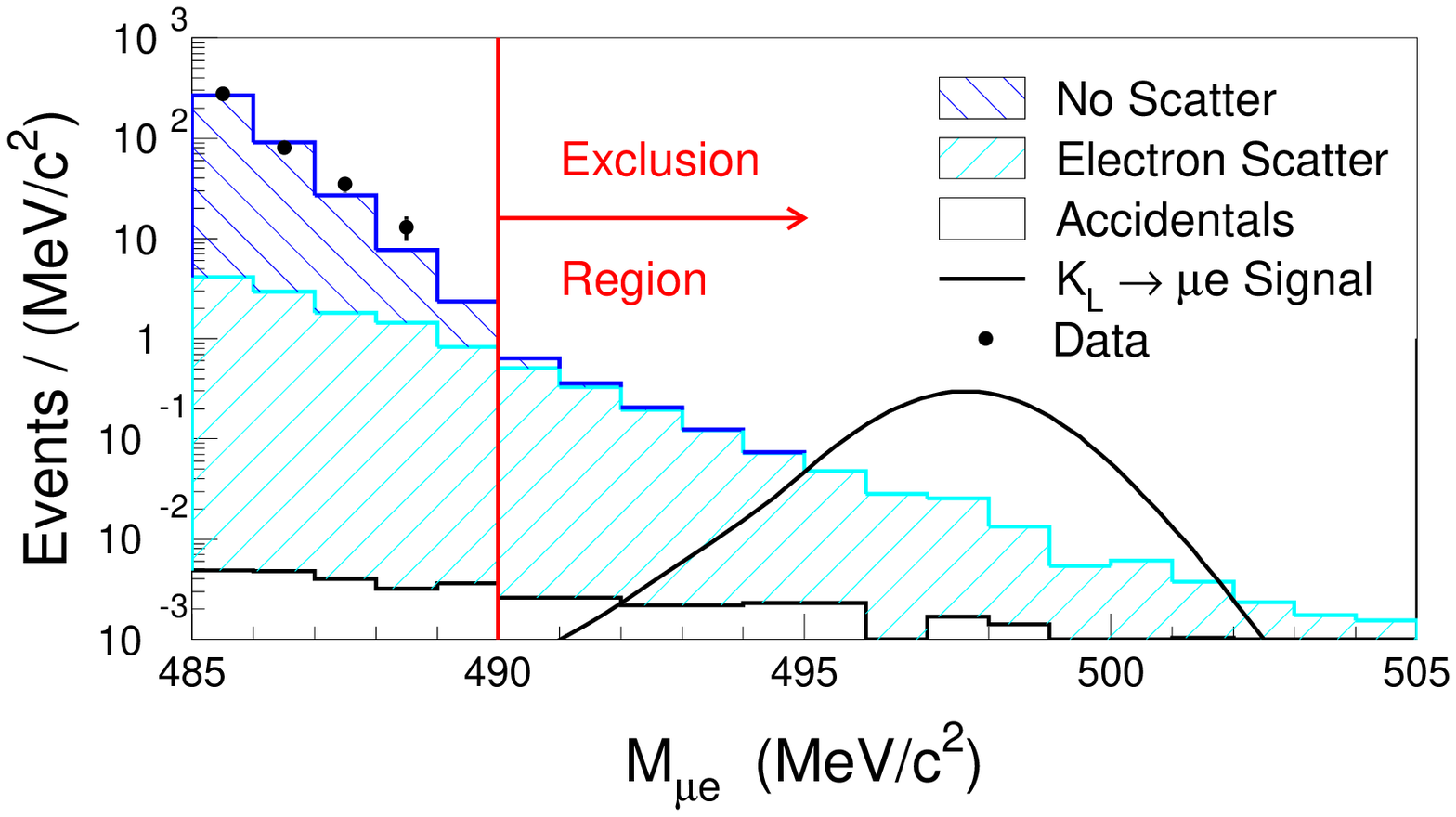,width=\linewidth}
\caption{Spectrum of $m_{\mu e}$ for events passing all other cuts (AGS E871).
The distribution of the three types of significant background are shown as
is the expected signal shape.
\label{fig:871mue2}}
\end{minipage}
\end{figure}

	AGS E871, the most sensitive $K$ decay experiment ever
mounted, was designed to detect $K_L \to e^{\pm}\mu^{\mp}$, 
$K_L \to e^+e^-,$ and $K_L \to \mu^+ \mu^-$.
The final result on $K_L \to \mu e$
was recently published~\cite{Ambrose:1998us}.  Fig.\ref{fig:871mue}
shows the distribution in $m_{\mu e}$ vs $p_T^2$ for events passing
all other cuts.  The rectangular contour bounds the region excluded
from cut optimization.  The inner contour bounds the actual signal
region. Fig.~\ref{fig:871mue2} shows the projection onto the $m_{\mu
e}$ axis of those events in Fig.~\ref{fig:871mue} with $p_T^2
<20$(MeV/c)$^2$.  Also shown are the calculated backgrounds, which
total $\le 0.1$ event in the signal region.  Since no events were
seen, a 90\% CL upper limit was extracted: $B(K_L \to \mu e)< 4.7
\times 10^{-12}$.  This is an 8-fold improvement on the previous limit
and corresponds to a remarkable mass scale of $\sim 150\,$TeV for a
hypothetical horizontal gauge boson (assuming standard electroweak
coupling strength).  This experiment was motivated largely by attempts
to explain the electroweak scale via dynamical symmetry breaking, and
results such as this one put this approach out of business.  The
experimenters believe that it would be possible to push their technique
another order of magnitude before the background due to Mott
scattering in chambers becomes intractable but there is no current
plan to continue this work.

\begin{figure}[h]
\begin{center}
\psfig{figure=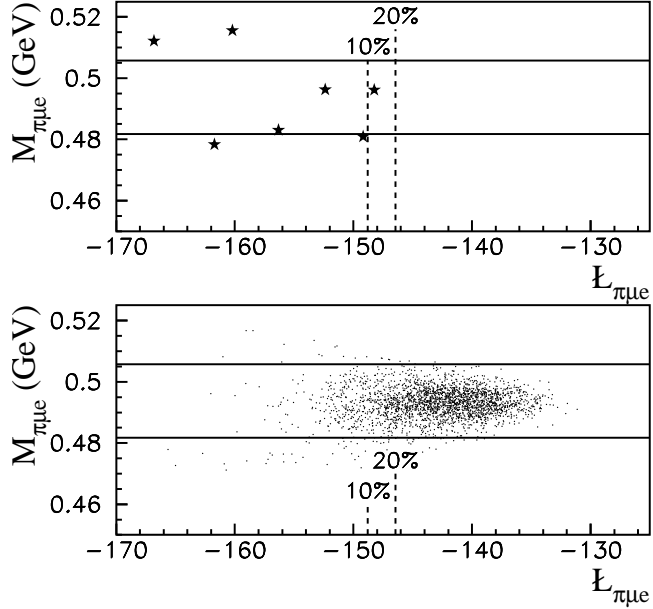,width=.5\linewidth}
\caption{Scatter plot of E865 1996 data (top) and Monte Carlo (bottom). The
horizontal lines demark the $3 \sigma$ mass region.
\label{fig:pme865}}
\end{center}
\end{figure}

	AGS E865 searched for the related decay $K^+ \to \pi^+ \mu^+
e^-$.  While $K_L \to \mu e$ is sensitive only to new pseudoscalar or
axial currents, the three body decay is sensitive to scalar or vector
currents. Moreover if LFV is observed in $K$ decay, the three body
decay will potentially be sensitive to details of the new interaction
through a study of the Dalitz Plot.  E865 has not completed analyzing
all its data.  However, they have recently released a result based on
the 1996 data~\cite{Appel:2000wg}.  Fig.~\ref{fig:pme865} is a
scatterplot of the log-likelihood of the reconstructed events under
the $K^+ \to \pi^+ \mu^+ e^-$ hypothesis versus the effective mass of
the detected particles.  A fit to the likelihood shapes of the signal
and background yields a 90\% CL upper limit of $B(K^+ \to \pi^+ \mu^+ e^-)
< 3.9 \times 10^{-11}$.  This is roughly five times better than the previous
limit, and when combined with previous results from this series of 
experiments yields  $B(K^+ \to \pi^+ \mu^+ e^-)< 2.8  \times 10^{-11}$.  A
further four-fold improvement in sensitivity is expected when all the
data presently on tape is analyzed. In addition, E865 has
new results~\cite{hong} on $K^+ \to \pi^+\mu^- e^+$ and $K^+ \to \pi^- \mu^+ e^+$,
as well as on $K^+ \to \pi^- \mu^+ \mu^+$ and $K^+ \to \pi^- e^+ e^+$.
This experiment also has many positive results on 
interesting kaon decays that I unfortunately don't have time to review.

Table~\ref{tab:lfv} summarizes the current state of searches for $K$ decays
that violate lepton flavor, lepton number, or both.  It's noticeable that even
the ``by-product'' results have in general reached sensitivities below $10^{-9}$.
Limits on some of these processes can be related to results on $\nu$-oscillation,
neutrinoless double $\beta$ decay and $\mu^- \to e^+$ conversion in the field of
a nucleus~\cite{Littenberg:2000fg}.

\begin{table}[t]
\caption{ \it Summary of searches for lepton flavor violating $K$ decays.}
\vskip 0.1 in
\begin{center}
\begin{tabular}{|l||l|l|l|l|} \hline
Decay Mode  &  \multicolumn{1}{|c|}{Branching Ratio} & Experiment &Pub. Date & Reference \\  \hline\hline
$K_L \to \mu e$  & $< 4.7\times10^{-12}$ & E871 &1998 &\cite{Ambrose:1998us}\\ \hline
$K^+ \to \pi^+ \mu^+ e^-$ & $< 2.8\times10^{-11}$ & E865 &2000&\cite{Appel:2000wg}\\ \hline
$K_L \to \pi^0\mu^{\pm} e^{\mp}$ & $< 4.4\times10^{-10}$  & KTeV &2000&\cite{ktevpme}\\ \hline
$K^+\rightarrow\pi^-\mu^+\mu^+$ & $<3\times10^{-9}$ & E865 &2000&\cite{hong}\\ \hline
$K^+\rightarrow\pi^- e^+ e^+$ & $<6.4\times10^{-10}$ & E865 &2000&\cite{hong}\\ \hline
$K^+\rightarrow\pi^+\mu^- e^+$ & $<5.2\times10^{-10}$ & E865 &2000&\cite{hong}\\ \hline
$K^+\rightarrow\pi^-\mu^+ e^+$ & $<5.0\times10^{-10}$ & E865 &2000&\cite{hong}\\ \hline
$K_L \to e^{\pm} e^{\pm} \mu^{\mp}\mu^{\mp}$ & $< 6.1\times10^{-9}$  & E799-I &1996&\cite{e799keemm}\\ \hline
\end{tabular}
\end{center}
\label{tab:lfv}
\end{table}

\section{One Loop Decays}

\begin{figure}[h]
\begin{center}
\psfig{figure=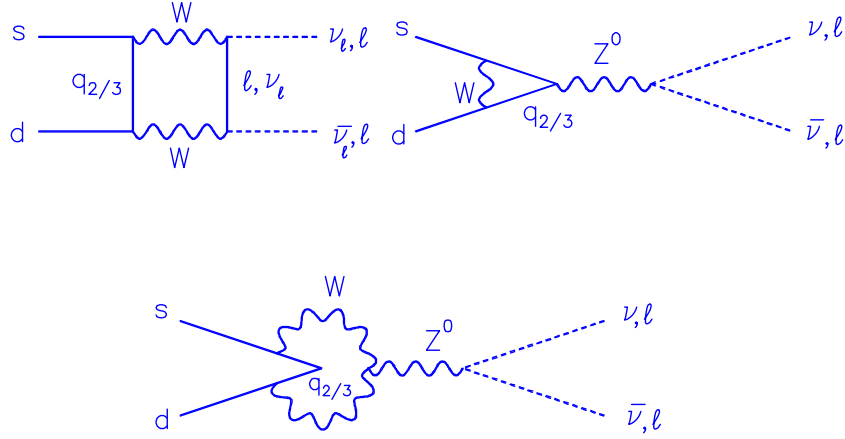,width=.65\linewidth}
\caption{One loop contributions to $K$ decay.
\label{fig:loops}}
\end{center}
\end{figure}

	As the activity in lepton flavor-violating kaon decays ramps
down, that in SM-allowed ``one-loop'' decays is ramping up.  These
GIM-suppressed~\cite{GIM} decays are dominated by, or at least have
measurable contributions from, loops involving weak bosons and heavy
quarks.  They include \kpnn0, \kpnnp, \kmm, $K_L \to \pi^0 e^+ e^-$
and $K_L \to \pi^0\mu^+\mu^-$.  In some cases such as \kpnn0, these
contributions violate CP.  The most interesting ones are those where
the loops dominate.  Fig~\ref{fig:loops} shows the Feynman diagrams
for these loops.  In general the processes are dominated by top quark
loops, which makes them directly sensitive to the quantity $\lambda_t
\equiv V^*_{ts}V_{td}$.  

In the past, it was common to assume CKM unitarity and discuss these
processes in terms of their sensitivity to $\rho$ and $\eta$ or 
$|V_{td}|$.  This is handy for comparison with information obtained from
the $B$ system, but it is perhaps unfair to $K$'s.  I will discuss an
alternative a little later, but for now I will keep to this convention.

\begin{figure}[h]
\begin{center}
\psfig{figure=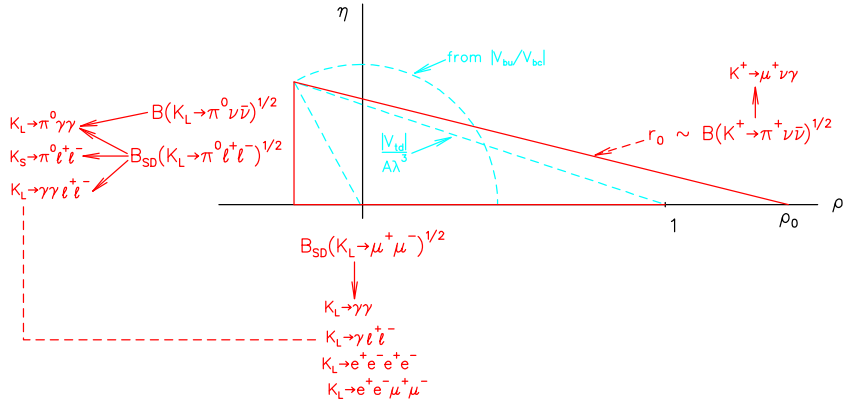,angle=90,width=.65\linewidth}
\caption{$K$ decays and the unitarity plane.
\label{fig:triangle}}
\end{center}
\end{figure}

	One can organize the CKM matrix information that can be obtained
from $K$ decays around the popular unitarity plane construction, as shown in
Fig.~\ref{fig:triangle}.  The lighter triangle is usual one, whereas the
darker triangle indicates the information available from rare kaon decays.
Note that the ``unitarity point'', $(\rho, \eta)$ is determined from
either triangle, and any disagreement between the $K$ and $B$ determinations
indicates physics beyond the SM.  The branching ratios closest to each line
can determine the length of that line.  The arrows from those branching ratios
point to processes that need to be studied either because they potentially
constitute backgrounds to the others, or because knowledge of them is useful
in connecting the branching ratio measurements with fundamental parameters.

\subsection{$K_L \to \mu^+ \mu^-$}

	$K_L \to \mu^+ \mu^-$ played a major role in the history of
weak interactions, where the observation of a surprisingly small rate
for this process was one of the inspirations for proposing the GIM
mechanism~\cite{GIM}.  There's a short distance contribution to the
amplitude that is rather reliably calculable in the SM~\cite{bandb}.
This contribution is proportional to the quantity $\rho_0 -
\rho$, where $\rho_0$ is a function of CKM $A$, $m_t$ and $m_c$,
and the QCD scale.  The QCD corrections to this amplitude have been
calculated to NLLA~\cite{bandb} and the residual uncertainty in $\rho_0$ 
due to this source is $< 10\%$.  Numerically $\rho_0
\approx 1.2$.  Thus a measurement of the rate for this process can
potentially determine $\rho$, as indicated in Fig.~\ref{fig:triangle}.  
Unfortunately, its usefulness in
this respect is limited by large long-distance
effects stemming from the $\gamma\gamma$ intermediate state.  This
is dominated by the absorptive contribution:
\bea
{\Gamma(K_L\to \gamma \gamma \to  \mu \mu)_{abs} \over \Gamma(K_L \to \gamma \gamma)}  =
\alpha^2 ({m_\mu \over m_K})^2 {1\over 2 \beta} {\left( \log{1+\beta \over 1-\beta} \right)}^2 = 1.195\times 10^{-5}
\eea\label{mmabs}
The resulting ``unitarity bound'' almost totally saturates the 
observed branching ratio.
Now this contribution can be calculated rather precisely, so that with
a sufficiently precise experiment, useful information on $\rho$ could
still be obtained.  Unfortunately there is also a dispersive 
contribution that is {\bf much} more difficult to calculate~\cite{mmld}
and that can interfere with the short-distance amplitude.

\begin{figure}[h]
\begin{center}
\psfig{figure=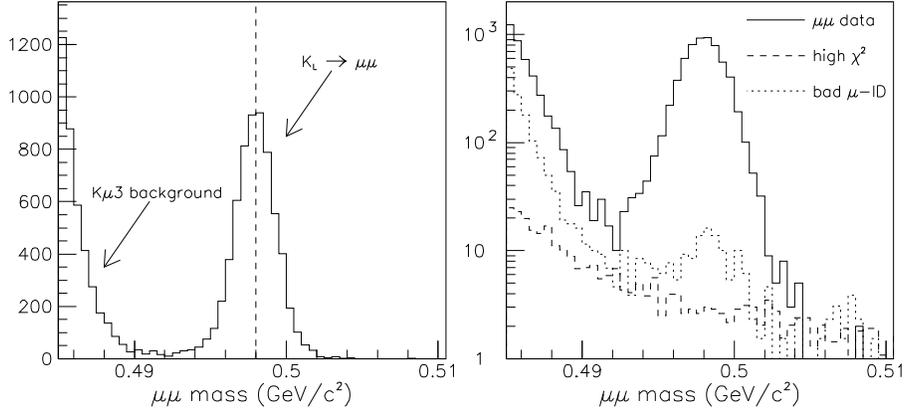,width=.70\linewidth}
\caption{Two body effective mass distribution for $K_L \to \mu^+ \mu^-$ candidates from AGS E871.
\label{fig:mm871}}
\end{center}
\end{figure}

The most recent study of this decay
is that of AGS E871 which observed a sample of approximately 6200 
events~\cite{mumu871}.  Fig.~\ref{fig:mm871} shows the $\mu \mu$ effective mass 
spectrum for $K_L \to \mu^+ \mu^-$ candidates.

	This data yields a branching ratio, $B(K_L \to \mu^+ \mu^-) =
(7.18 \pm 0.17)\times 10^{-9}$, which is to be compared with the
unitarity bound from Eq.~\ref{mmabs} of $(7.07 \pm 0.18)\times
10^{-9}$.  The difference yields a 90\% CL upper limit on the total
dispersive contribution to $(K_L \to \mu^+ \mu^-)$ of $0.37 \times
10^{-9}$.  The E871 collaborators attempted to properly take the long
distance contribution to the dispersive amplitude into account in
extracting $\rho > -0.33$ at 90\% CL from this limit.  I will take a
somewhat different approach below in exploiting this result.
One can't leave the discussion of this experiment without mentioning
the tour-de-force measurement~\cite{ee871}, $B(K_L \to e^+ e^-) = 
(8.7 {+5.7 \atop{-4.1}}) \times 10^{-12}$, which is the smallest particle 
branching ratio ever reported. 

	There is no near-term plan for another experimental study of $K_L \to
\mu^+ \mu^-$.  Further progress in the extraction of $\rho$ will depend on
developments in theory.  However it is thought that these developments
can be advanced by the results of experiments on processes of the form
$K_L \to \gamma \gamma$ in which one or both of the gammas is virtual.
This hope is easy to understand, since it is off-shell intermediate states
that contribute to the dispersive amplitude for $K_L \to
\mu^+\mu^-$.  The processes involved include 
(1) $K_L \to \gamma e^+ e^-$, (2) $K_L \to \gamma \mu^+ \mu^-$, 
(3) $K_L \to e^+ e^- e^+ e^-$, (4)  $K_L \to e^+ e^- \mu^+ \mu^-$, 
and, in principle (5) $K_L \to \mu^+ \mu^- \mu^+ \mu^-$.  
There has been recent data on (1)~\cite{eeg}, (2)~\cite{mmg}, (3)~\cite{mmg},
and (4)~\cite{mmg}.  I would say that the jury is still out on how well
theory copes with the dispersive amplitude.

\subsection{\kpnnp}

\begin{figure}[h]
\begin{minipage}[b]{.445\linewidth}
\psfig{figure=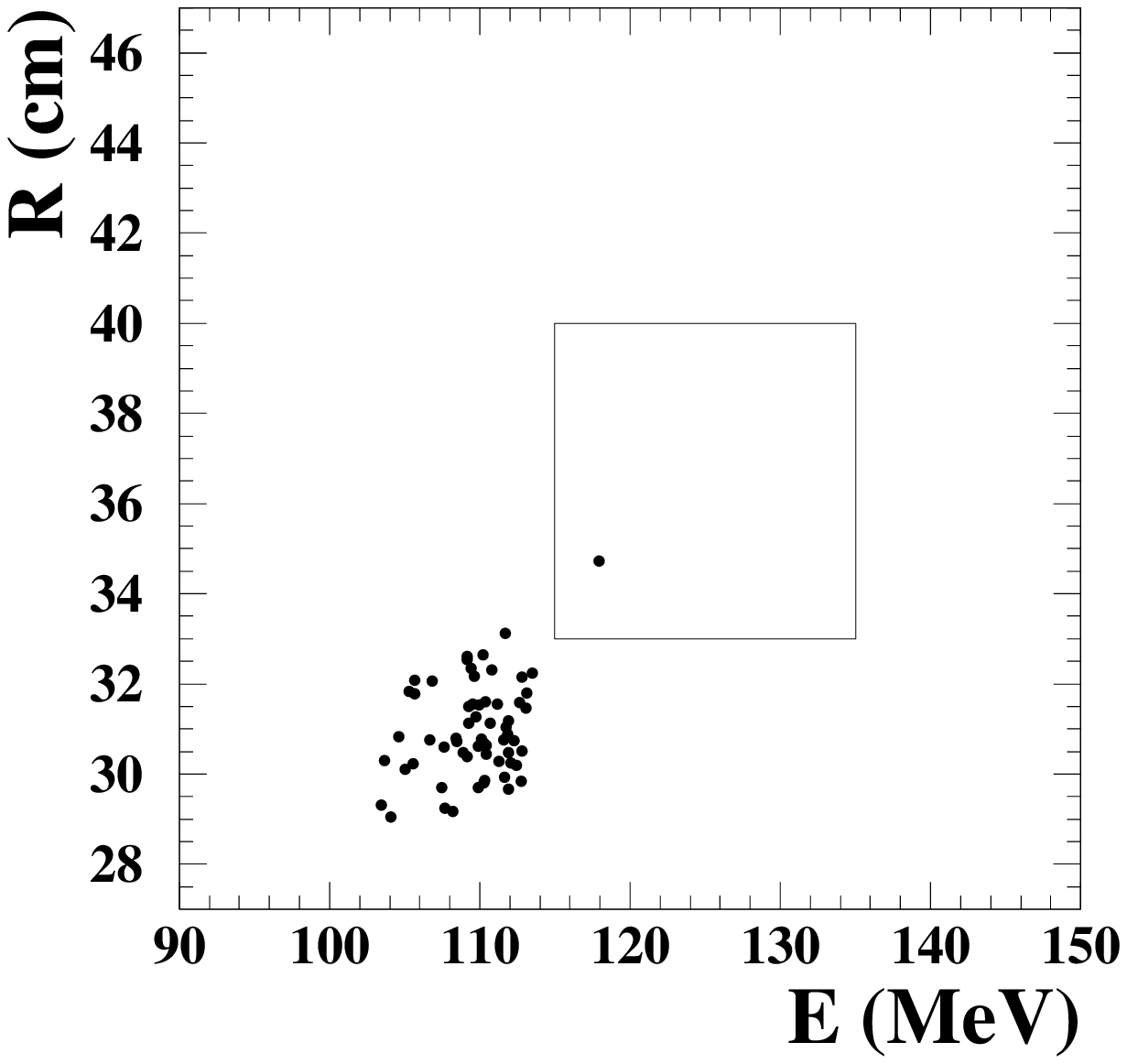,width=\linewidth}
\caption{The $\pi^+$ range vs kinetic energy for events passing all other cuts 
from AGS E787.  Box indicates \kpnnp\ signal region.  Events at lower left
are residual $K^+ \to \pi^+ \pi^0$.
\label{fig:pnnrslt}}
\end{minipage}\hfill
\begin{minipage}[b]{.475\linewidth}
\psfig{figure=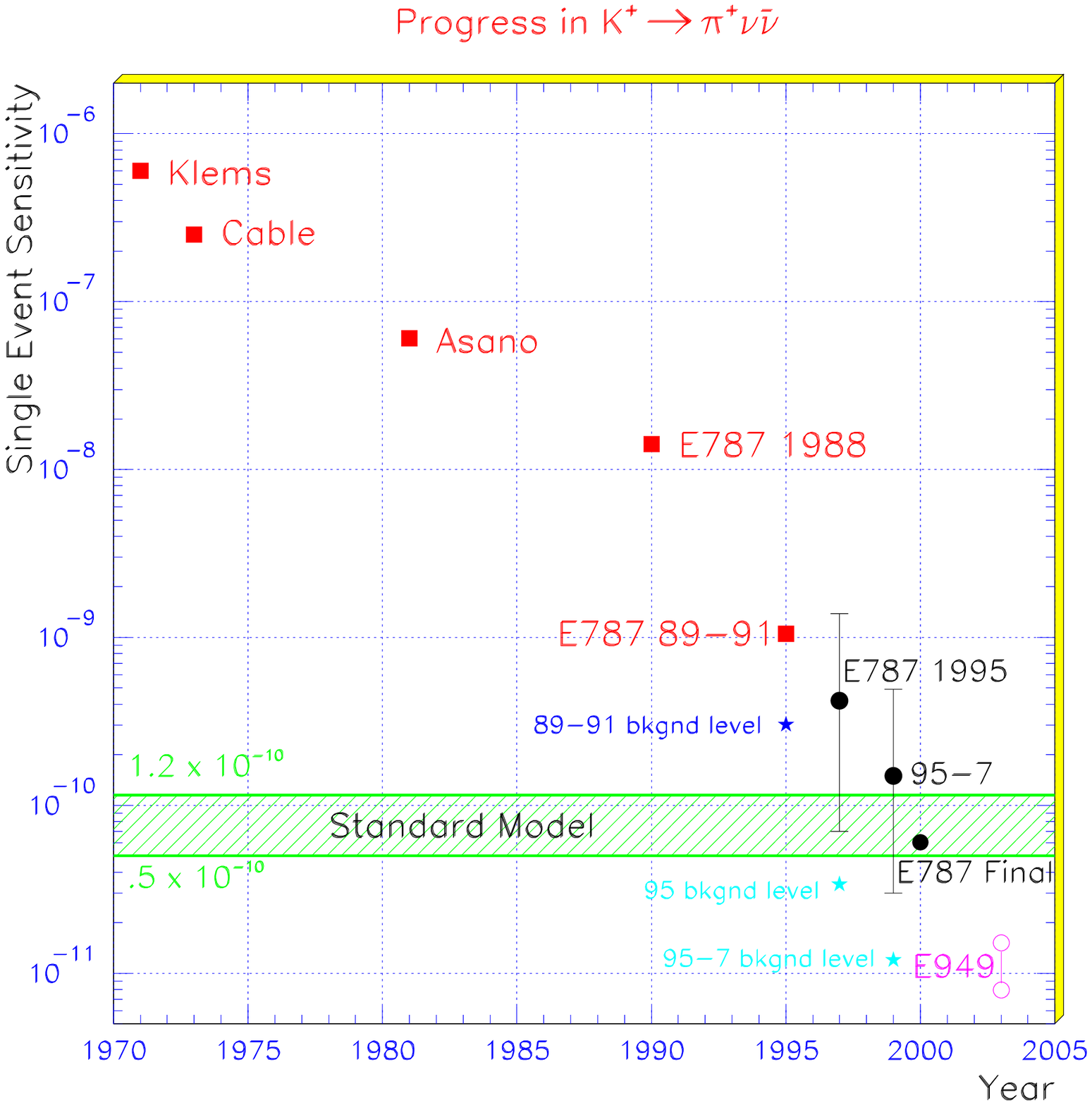,width=\linewidth}
\caption{The history and projected future progress of the study of \kpnnp\ 
compared to Standard Model expectation.  AGS E949 will extend this study to 
the $10^{-11}$/event level by around 2004.  
\label{fig:prog}}
\end{minipage}
\end{figure}

	\kpnnp\ is another process
sensitive to short distance physics, that has none of the theoretical
problems of $K_L \to \mu^+\mu^-$, but unfortunately is much more
difficult to study.  There are no significant long-distance contributions
to this decay, and the usually problematic hadronic matrix element can
be calculated via an isospin transformation from that of the well-measured
$K_{e3}$ decay~\cite{2mar}.  This decay is usually discussed in terms of its
sensitivity to $V_{td}$.  The amplitude for this decay is proportional to
the dark slanted line at the right in Fig.~\ref{fig:triangle}.  This is
equal to the vector sum of the line proportional to $|V_{td}|/A \lambda^3$
and that from 1 to the point marked $\rho_0$. This length along the real 
axis is proportional to amplitude for the charm contribution to 
\kpnnp.  The QCD corrections to this amplitude
are the source of the largest theoretical uncertainty in the calculation of 
\bkpnnp.  These have been calculated to NLLA~\cite{bandb}, and the residual 
uncertainty in the charm amplitude is estimated to be $\sim 15\%$.  This 
results in a $\sim 6\%$ uncertainty~\cite{Buchalla:1999ba} in
extracting $|V_{td}|$ from \bkpnnp.  Experiment is
still far from this level.  In 1997, AGS E787 published evidence for
one event of \kpnnp~\cite{Adler:1997am}.  Recently an analysis of a larger
data set has been published~\cite{Adler:2000by}.  Fig.~\ref{fig:pnnrslt}
shows the results: no further events were seen, leading to a branching 
ratio \bkpnnp $= (1.5 {+3.4 \atop -1.2}) \times 10^{-10}$. This is
to be compared to the expectation of $(0.82 \pm 0.32) \times 10^{-10}$ from 
fits to the CKM phenomenology~\cite{Buchalla:1999ba}.  E787 has a further
sample under analysis of sensitivity about equal to the sum of all its 
previous runs.  It is notable that E787 has established methods to reduce the 
residual background to $\sim 10\%$ of the the signal branching ratio 
predicted by the SM.

A new experiment, AGS E949~\cite{e949}, based on an upgrade of the E787
detector, is in preparation and scheduled to run in 2001-3.  Using
the entire flux of the AGS, it is expected
to reach a sensitivity of $\sim 10^{-11}$/event.  Fig.~\ref{fig:prog} shows
the history and near term expectations of the progress in studying this decay.
All the experiments in Fig.~\ref{fig:prog} used stopped-$K^+$ beams.
In the longer term, the CKM experiment~\cite{CKM} at Fermilab proposes
to use an in-flight technique to reach $10^{-12}$/event sensitivity.
If CKM fits to the SM phenomenology are correct, this will result in some
70 \kpnnp\ events.

\subsection{\kpnn0}

	\kpnn0 is the holy grail of the $K$ system.  It is direct CP-violating
to a very good approximation~\cite{ll89,buchalla}, like \kpnnp\ it has a 
hadronic matrix element which can be determined from that of $K_{e3}$, but, 
unlike \kpnnp, it has no significant contribution from charm.  Thus the
intrinsic theoretical uncertainty on the connection between \bkpnn0\ and
the fundamental SM parameters is only about 2\% \footnote{Note \bkpnn0$\propto A^4 
\eta^2$.}. 

	Our best knowledge of \bkpnn0\ comes indirectly from E787's measurement
of \bkpnnp\ through a model independent relationship pointed out
by Grossman and Nir~\cite{Grossman:1997sk}: \bkpnn0$< 4.4\, $\bkpnnp.  In the
SM this is equivalent to the statement that the imaginary part of an
amplitude cannot be larger than its modulus.  The E787 result cited
above then yields \bkpnn0$<2.6 \times 10^{-9}$ at 90\% CL.  This is much
tighter than the current direct experimental limit, $5.9 \times 10^{-7}$,
which comes from the KTeV experiment at Fermilab~\cite{Alavi-Harati:2000hd}.
However to actually measure \bkpnn0, it will be necessary to improve
the reach of direct experiment by some five orders of magnitude.  The
KEK E391a experiment~\cite{e391a} does not quite propose to bridge this gap, 
but to achieve a sensitivity of $\sim 10^{-10}$/event, which would at least 
better the indirect limit by an order of magnitude.  It will serve as a
test for a future much more sensitive experiment to be performed at
the Japanese Hadron Facility.  

	At Fermilab, the KaMI~\cite{kami} proponents plan to use the high
proton current of the Main Injector to make a $K_L$ beam
sufficiently intense that a sensitivity of $<10^{-12}$/event
for \kpnn0\ can, in principle, be achieved.  This experiment is
similar to KEK E391a, with the major exception that the energy scale is a
factor of ten higher.  Both feature a pencil beam, a crystal
spectrometer and an hermetic photon veto, all operating within an
evacuated enclosure. KaMI would also incorporate a scintillating fiber
charged particle spectrometer, greatly enhancing the
physics menu of the experiment.

	A completely different approach is taken by the KOPIO
experiment~\cite{e926} which will exploit the intensity and flexibility of
the BNL AGS to make a high-flux, low-energy, microbunched $K_L$ beam.
This allows time-of-flight determination of the $K_L$ velocity.  In
addition, the direction as well as energy of the final state photons
will be measured, so that a well-defined vertex can be found. This provides
a measurement of the $K_L$ 3-momentum so that
kinematic constraints as well as photon vetoing are available
to fight backgrounds.  The leading expected background
is \kpp0, which is some eight orders of magnitude larger than the
predicted signal.  Since $\pi^0$'s from \kpp0\ have a unique energy in
the $K_L$ center of mass, a very effective kinematic cut can be
applied.  This reduces the burden on the photon veto system to the
point where the techniques proven in E787 are sufficient.  This
experiment will avoid having to operate most of its apparatus in a vacuum
at the cost of having a thin vacuum enclosure around the beam.  KOPIO aims
to collect 65 \kpnn0\ events with a signal to background
ratio of 2:1.  This will permit $\eta$ to be determined to $<10\%$,
given expected progress in measuring $m_t$ and $V_{cb}$.  KOPIO will
run during the $\sim$20 hours/day the AGS is
not needed for injection into RHIC.

\subsection{A different way of using rare kaon decay results}

	Many theorists advocate comparing the
results of rare kaon decay experiments with those from the $B$ system
when both are available.  Obviously $|V_{td}|$ derived from \bkpnnp\ can be
checked against $|V_{td}|$ extracted from the ratio of $\bar B_s-B_s$ and $\bar
B_d-B_d$ mixing.  It has also been emphasized that results from \kpnnp\
and \kpnn0\ can be combined to yield
$sin 2\beta$~\cite{Buchalla:1994tr,Buchalla:1996fp} or a quantity closely 
related to it~\cite{Grossman:1997sk,Nir:1998tf,Bergmann:2000ak}.
A number of other useful relations have also 
been pointed out~\cite{Buchalla:1999ba,Buchalla:1996fp}.

	However, I favor a different approach, in which 
$\rho$ and
$\eta$ are de-emphasized in favor of the real and imaginary part of
$\lambda_t$.  $Im(\lambda_t)$ is closely
related to the Jarlskog invariant, $\cal{J_{CP}}$~\cite{ja}, and thus
to the area of the unitarity triangle,
\bea
{\cal{J_{CP}}} = Im(V_{ud}^* V_{us} V_{ts}^* V_{td}) = 
\lambda(1 - \frac{\lambda^2}{2} ) Im(\lambda_t)
\eea\label{eq:jarlsk}
In this approach, the branching ratio for \kpnn0\ can be written:
\begin{equation}\label{br}
B(K_L \to \pi^0 \nu\bar\nu) = {{3 \tau_{L} \alpha^2 r_{L} B_{K^+e3}}\over{\tau_{+} V_{us}^2 2 \pi^2 sin^4{\theta_W}}} (Im \lambda_t X(x_t))^2
\end{equation}
where we take $\alpha=1/129$ and $sin^2{\theta_W} = 0.23$, 
$r_{L}$ is an isospin-breaking and
phase space correction~\cite{2mar},
which for this process $= 0.944$.  $X(x_t)$ is an
Inami-Lim~\cite{inami} function of $x_t \equiv m_t^2/m_W^2$ which
$\approx 0.65 x_t^{0.59}$.
The product $r_{L} B_{K^+e3}$ accounts for the hadronic matrix element,
and the $\tau$'s are the $K_L$ and $K^+$ lifetimes.
Since all these factors are well determined,\footnote{Already the {\it current}
uncertainty on $m_t$ would lead to only a 3.5\% contribution to the 
uncertainty in 
$Im(\lambda_t)$ measured from \bkpnn0.} a measurement of \bkpnn0\ gives
a direct measurement of the area of the unitarity triangle.  This
can then be compared with any of a number of different indirect determinations
of the unitarity triangle area from studies of $B$'s.

	It is instructive to write the branching ratio for \kpnnp\ in a
form equivalent to Eq.~\ref{br}:
\begin{equation}\label{brp}
B(K^+ \to \pi^+ \nu\bar\nu) = {{r_{+} \alpha^2 B_{K^+e3}}\over{V_{us}^2 2 \pi^2 sin^
4{\theta_W}}}
\sum_i |\lambda_c X^i_{NL} + \lambda_t X(x_t)|^2
\end{equation}
where $r_{+} = 0.901$, $\lambda_c \equiv V_{cs}^* V_{cd}$, 
$i = e, \mu, \tau$, and $X_{NL}^i$ is given in 
Table 1 of Ref.~\cite{Buchalla:1999ba}.

	Now whereas there are no useful limits on $\lambda_t$ from
Eq.~\ref{br}, one can use Eq.~\ref{brp} to extract limits from the
latest result of E787.  One can obtain~\cite{Adler:2000by}:
\begin{eqnarray}
~~~~~~~~& ~~~~~| Im(\lambda_t)|& < 1.22 \times 10^{-3}\label{results1}\\
-1.10 \times 10^{-3}&<~~~  Re(\lambda_t)~~~~~~~& < 1.39 \times 10^{-3} \label{results2}\\
 1.07 \times 10^{-4}& <~~~ |\lambda_t|~~~~~~~& < 1.39 \times 10^{-3}.\label{results3}
\end{eqnarray}
Note that the actual regions of the $\lambda_t$ plane constrained by \kpnnp\
are (approximately) annuli to which the above one-dimensional limits are 
tangent.  The charm contribution displaces the center of the limiting
circles from the origin. To get conservative ``outer'' limits one must 
minimize $m_t$.

One can write down a similar form for the short-distance part of
the \kmm\ branching ratio.
\begin{eqnarray}\label{brmm}
B^{SD}(K_L \to \mu\mu)& = &{ {\tau_{K_L} \alpha^2 B_{K^+ \mu \nu}}\over{\tau_{K^+} 
V_{us}^2 \pi^2 sin^4{\theta_W}}} [Re(\lambda_c) Y_{NL} + Re(\lambda_t) Y(x_t)]^2
\end{eqnarray}
where $Y(x_t) \approx 0.32 x_t^{0.78}$
and $Y_{NL}$ is given in Table 3 of Ref.~\cite{Buchalla:1999ba}.
To make use of the current experimental result on \bkmm, one must
face the problem of the long-distance dispersive amplitude. A
choice that has sometimes been made in the experimental literature is
to limit the possible absolute size of this amplitude by 
the 90\% CL limit of a recent calculation~\cite{D'Ambrosio:1998jp}:
$|Re A_{LD}| < 2.9 \times 10^{-5}$.   With some trepidation
I use it, partly on the ground that considerably less conservative 
assumptions have sometimes been made in the same literature.
For example, in extracting the limit on $\rho$ mentioned above,
the E871 collaboration used the probability distribution of 
$|Re A_{LD}|$ given in Ref.~\cite{D'Ambrosio:1998jp}, rather than imposing the
90\% CL limit~\cite{ambrose}.  Having made the more conservative choice, and 
taking $m_t$ and $Y_{NL}$ to the limits recommended in 
Ref.~\cite{Buchalla:1999ba} I obtain 
\begin{equation}\label{mmresult}
-5.85 \times 10^{-4} < Re(\lambda_t) < 7.24 \times 10^{-4}
\end{equation}

\begin{figure}[h]
\begin{minipage}[b]{.445\linewidth}
\psfig{figure=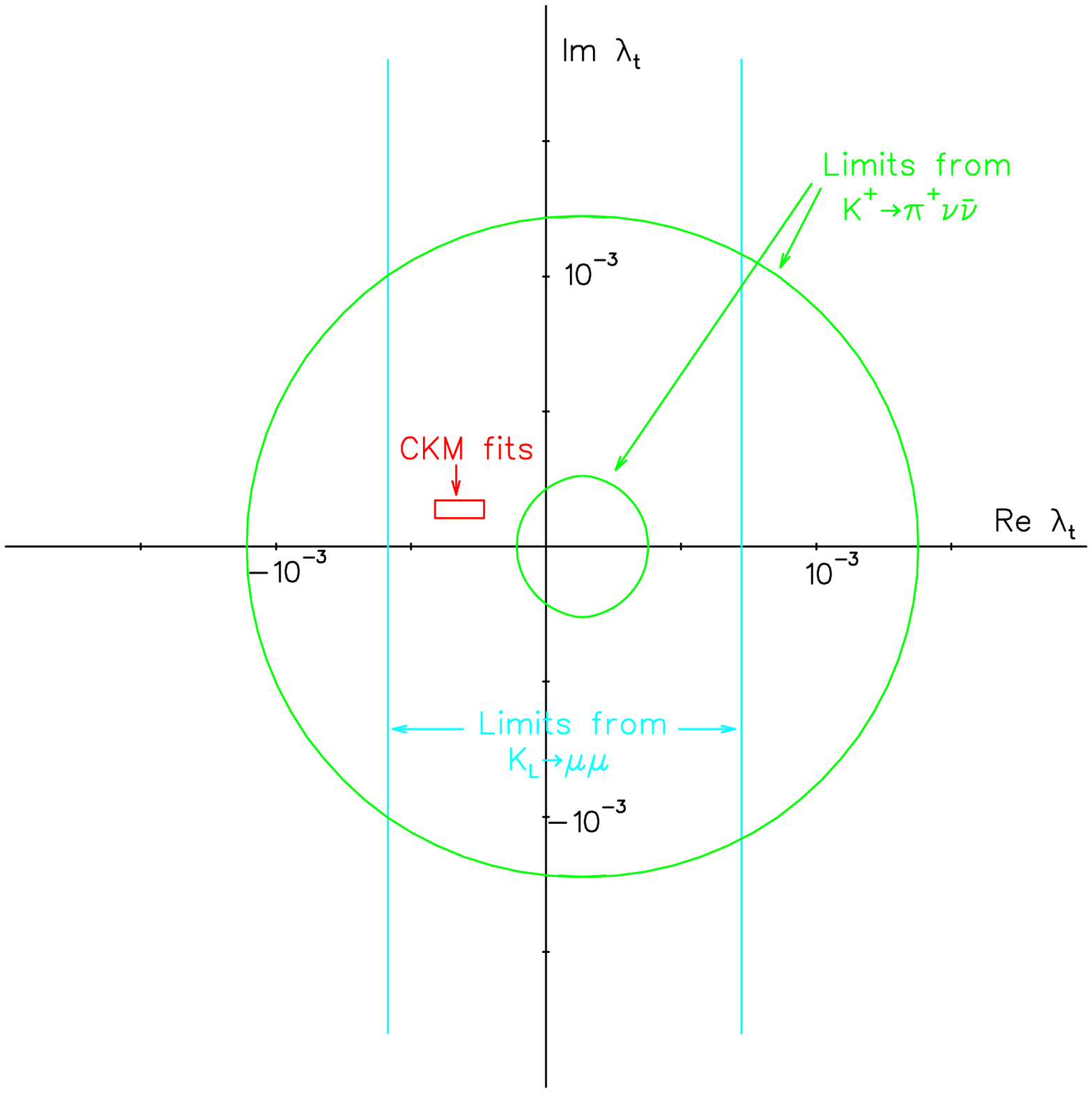,width=\linewidth}
\caption{Region allowed by the two rare K decays, compared with current 
CKM fits.
\label{fig:present}}
\end{minipage}\hfill
\begin{minipage}[b]{.405\linewidth}
\psfig{figure=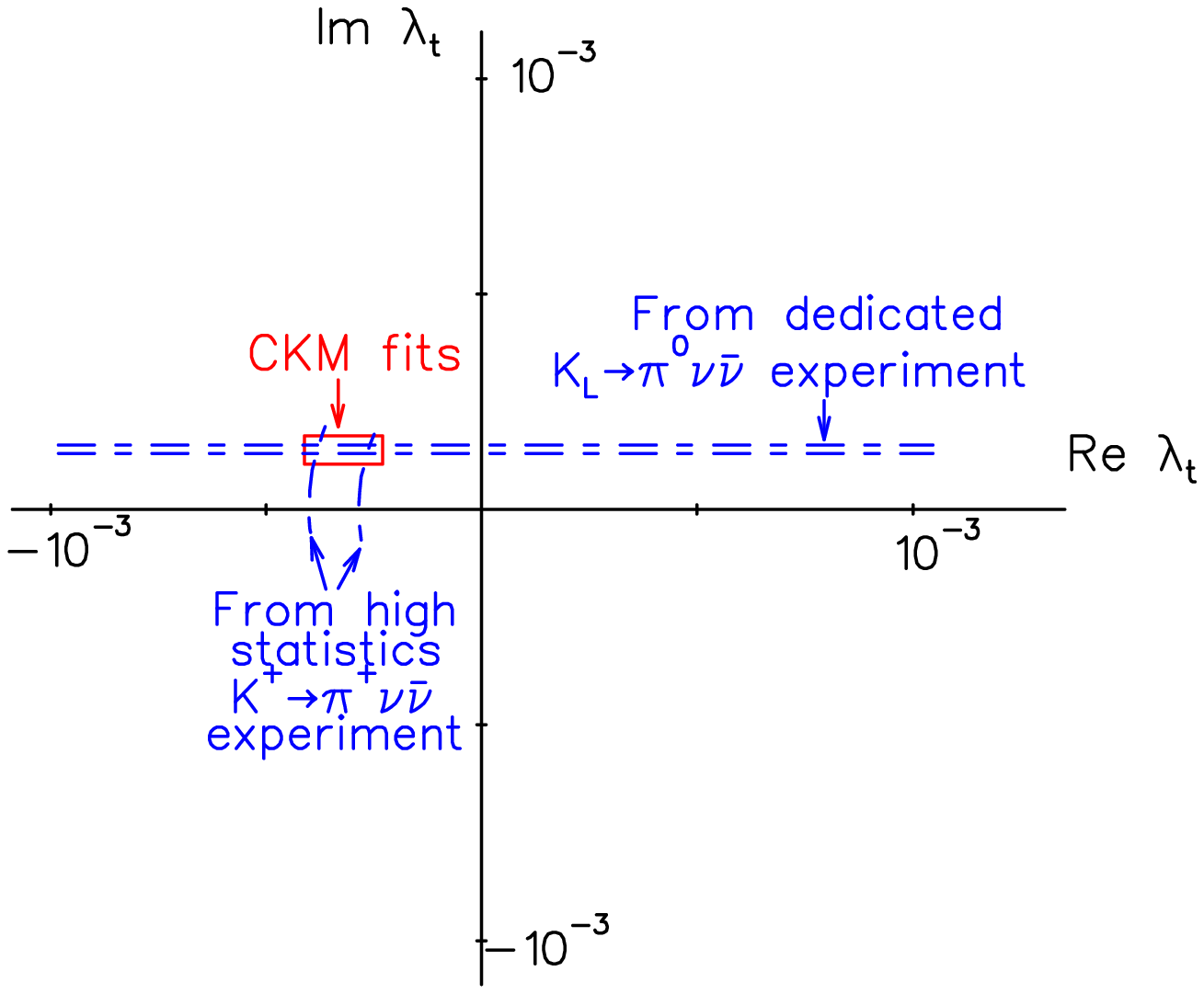,width=\linewidth}
\caption{Comparison between present CKM phenomenology and results from
future experiments on charged and neutral $K \to \pi \nu\bar\nu$.
\label{fig:proj}}
\end{minipage}
\end{figure}

Fig.~\ref{fig:present} shows the region of the $\lambda_t$ plane constrained
by \kpnnp\ and \kmm, compared to results from recent CKM fits~\cite{bands}.
The allowed region is that both between the circles and between the 
vertical lines.
These limits are weaker than those that can be obtained from the full
available CKM  phenomenology
but they are independent of information obtained from the $B$ system, or
from CP-violation measurements in $K \to \pi\pi$, thus the observed 
agreement is quite meaningful.

\subsection{Conclusions}

Although I couldn't cover much of this, the kaon decay
sections of the Particle Data Book are being largely rewritten by the 
results of current and recent experiments.  

Lepton flavor violation experiments have been pushed to
remarkable sensitivities, corresponding to mass scales of well over 100
TeV.  But this success has killed most models predicting LFV in kaon decay.
Barring developments in theory, this subject will probably advance only
as a by-product of other studies.

	The recent precision measurement of \kmm\ will be very useful if
theorists can nail down the size of the dispersive long-distance amplitude.

	\kpnnp\ has been seen and can clearly be further exploited.
There are two coordinated initiatives devoted to this: a
$10^{-11}$/event experiment is being prepared at the BNL AGS
and a $10^{-12}$/event experiment at the FNAL Main Injector is in the 
R\&D phase.  The first dedicated experiment to
seek \kpnn0\ is proceeding and two other initiatives are in progress
with the goal of making a $\sim 10\%$ measurement of $Im(\lambda_t)$.

	The motivation for pursuing $K \to \pi \nu\bar\nu$ is stronger
than ever.  An ``alternative'' unitarity triangle can be constructed from
this data, and it will be invaluable for comparison with results from
the $B$ system if new physics is at work in the flavor sector.  
By and large the effects of such new physics will be
quite different in the $K$ and $B$ systems.  Fig.~\ref{fig:proj} shows
what $K \to \pi\nu\bar\nu$ experiments might offer in the next few
years.  This figure assumes the two sectors will agree, but it is important
to notice the large area available for them {\it not} to do so!

\section*{Acknowledgments}
I thank D. Ambrose, D. Bryman, A.J.Buras, M. Diwan, S. Kettell, W.J.Marciano, 
W. Molzon, J. Ritchie, and M. Zeller for useful discussions, access to 
results, and other materials.
This work was supported by the U.S.  Department of Energy under
Contract No. DE-AC02-98CH10886.


\section*{References}
\begin{thebibliography}{99}
\bibitem{GIM}S.L. Glashow, J. Iliopoulos and L. Maiani, \Journal{\PRD}{2}{1285}
{1970}.

\bibitem{steve}A.~R.~Barker and S.~H.~Kettell,
hep-ex/0009024.

\bibitem{Ambrose:1998us}D.~Ambrose {\it et al.}  [E871 Collaboration],
\Journal{\PRL}{81}{5734}{1998}.

\bibitem{Appel:2000wg}R.~Appel {\it et al.},
\Journal{\PRL}{85}{2450}{2000}
hep-ex/0005016.

\bibitem{hong}R. Appel {\it et al.}, \Journal{\PRL}{85}{2877}{2000}
hep-ex/0006003.

\bibitem{ktevpme}A. Bellavance, Proc. Meet. DPF, Columbus OH, August 2000
Singapore: (World Sci. 2001).

\bibitem{Littenberg:2000fg}L.~S.~Littenberg and R.~Shrock,
hep-ph/0005285.

\bibitem{e799keemm}P. Gu {\it et al.}, \Journal{PRL}{76}{4312}{1996}.

\bibitem{bandb}G.~Buchalla and A.~J.~Buras, \Journal{\NPB}{412}{106}{1994}.

\bibitem{mmld}G.~Valencia, \Journal{\NPB}{517}{339}{1998};
G.~D'Ambrosio, G.~Isidori, and J.~Portol\'es, \Journal{\NPB}{423}{385}{1998}; 
D.~Gomez-Dumm and A.~Pich \Journal{\PRL}{80}{463}{1998};
M.~Knecht, S.~Peris, M.~Perrottet and E.~de Rafael
\Journal{\PRL}{83}{5230}{1999}.

\bibitem{mumu871}D.~Ambrose {\it et al.}  [E871 Collaboration],
\Journal{\PRL}{84}{1389}{2000}.

\bibitem{ee871}D.~Ambrose {\it et al.}  [E871 Collaboration],
\Journal{\PRL}{81}{4309}{1998}.

\bibitem{eeg}V.~Fanti {\it et al.}, \Journal{\PLB}{458}{553}{1999}.

\bibitem{mmg}Y.~Wah, ICHEP (2000).

\bibitem{2mar}W. Marciano and Z. Parsa, \Journal{\PRD}{53}{1}{1996}.

\bibitem{Buchalla:1999ba}
G.~Buchalla and A.~J.~Buras, \Journal{\NPB}{548}{39}{1999}.
[hep-ph/9901288].

\bibitem{Adler:1997am}
S.~Adler {\it et al.}  [E787 Collaboration],
\Journal{\PRL}{79}{2204}{1997}.

\bibitem{Adler:2000by}
S.~Adler {\it et al.}  [E787 Collaboration],
\Journal{\PRL}{84}{3768}{2000}.

\bibitem{e949}B. Bassalleck {\it et al}., E949 Proposal, BNL-67247,
TRI-PP-00-06, August 1999.

\bibitem{CKM}R. Coleman, {\it et al.}, ``Charged Kaons at the Main 
Injector'', FNAL proposal, April 15, 1998, FERMILAB-P-0905.


\bibitem{ll89}L.S.~Littenberg, \Journal{\PRD}{39}{3322}{1989}.

\bibitem{buchalla}G.~Buchalla and G.~Isidori, \Journal{\PLB}{440}{170}{2000}.

\bibitem{Grossman:1997sk}
Y.~Grossman and Y.~Nir, \Journal{\PLB}{398}{163}{1997}.

\bibitem{Alavi-Harati:2000hd}
A.~Alavi-Harati {\it et al.}, \Journal{\PRD}{61}{072006}{2000}.

\bibitem{e391a} T.Inagaki {\it et. al.}, KEK Internal 96-13, November 1996.

\bibitem{kami} E.Chen {\it et. al.}, hep-ex/9709026.

\bibitem{e926} I.-H.~Chiang {\it et al.}, AGS Experiment Proposal 926 (1996).

\bibitem{Buchalla:1994tr}G.~Buchalla and A.J.~Buras,
\Journal{\PLB}{333}{221}{1994}.

\bibitem{Buchalla:1996fp}
G.~Buchalla and A.~J.~Buras,
\Journal{PRD}{54}{6782}{1996}.

\bibitem{Nir:1998tf}
Y.~Nir and M.~P.~Worah,
\Journal{\PLB}{423}{319}{1998}.

\bibitem{Bergmann:2000ak} S.~Bergmann and G.~Perez,
JHEP {\bf 0008}, 034 (2000).

\bibitem{ja}
C.~Jarlskog, Phys. Rev. Lett. {\bf 55}, 1039 (1985).

\bibitem{inami}T.~Inami and C.S.~Lim, {\it Prog. Theor. Phys.}, {\bf 65},
297 (1981), Erratum-{\it ibid.} {\bf 65}, 1772 (1981).

\bibitem{D'Ambrosio:1998jp}
G.~D'Ambrosio, G.~Isidori and J.~Portoles,
\Journal{\PLB}{423}{385}{1998}.

\bibitem{ambrose}D. Ambrose, private communication.

\bibitem{bands} A.J.~Buras and L.~Silvestrini, \Journal{\NPB}{546}{299}{2000}.

\end{thebibliography}
\end{document}